\documentclass[twocolumn]{aastex63}

\newcommand{\Msun}{\,\text{M}_{\odot}}

\begin{document}

\title{$R$-mode Stability of GW190814's Secondary Component as a Supermassive and Superfast Pulsar}

\correspondingauthor{Ang Li}
\email{liang@xmu.edu.cn}

\author{Xia Zhou}
\affiliation{Xinjiang Astronomical Observatory, Chinese Academy of Sciences, Urumqi, Xinjiang 830011, China}

\author{Ang Li}
\affiliation{Department of Astronomy, Xiamen University, Xiamen, Fujian 361005, China}

\author{Bao-An Li}
\affiliation{Department of Physics and Astronomy, Texas A\&M University-Commerce, Commerce, TX 75429, USA}

\date{\today}

\begin{abstract}
The nature of GW190814's secondary component $m_2$ of mass $(2.50-2.67)\Msun$ in the mass gap between the currently known maximum mass of neutron stars and the minimum mass of black holes is currently under hot debate. Among the many possibilities proposed in the literature, the $m_2$ was suggested as a superfast pulsar while its r-mode stability against the run-away gravitational radiation through the Chandrasekhar-Friedman-Schutz mechanism is still unknown. Using those fulfilling all currently known astrophysical and nuclear physics constraints among a sample of 33 unified equation of states (EOSs) constructed previously by Fortin {\it et al.} (2016) using the same nuclear interactions from the crust to the core consistently, we compare the minimum frequency required for the $m_2$ to rotationally sustain a mass higher than $2.50\Msun$ with the critical frequency above which the r-mode instability occurs. We use two extreme damping models assuming the crust is either perfectly rigid or elastic. Using the stability of 19 observed low-mass x-ray binaries as an indication that the rigid crust damping of the r-mode dominates within the models studied, we find that the $m_2$ is r-mode stable while rotating with a frequency higher than 870.2 Hz (0.744 times its Kepler frequency of 1169.6 Hz) as long as its temperate is lower than about $3.9\times 10^7 K$, further supporting the proposal that GW190814's secondary component is a supermassive and superfast pulsar.
\end{abstract}

\keywords{
Pulsars (1306);
Neutron stars (1108)
}
%
\section{Introduction}
The recent discovery by the LIGO/Virgo Collaborations that the compact binary merger GW190814 has a secondary component $m_2$ with mass $(2.50-2.67)\Msun$~\citep{2020ApJ...896L..44A} at $90\%$ credible level has generated much interest in the astrophysics community. As the $m_2$ is in the mass gap between the currently known maximum mass of neutron stars and the minimum mass of black holes, its nature has significant ramifications on many interesting issues in astrophysics and cosmology. Many possibilities regarding its nature and formation path have been proposed very recently in the literature; see, e.g.,~\citet{2020arXiv201011020R,2020arXiv201002090B,2020arXiv201001509B} among the latest reports and references therein.

Since the maximum mass $M_{\rm TOV}$ of non-rotating neutron stars is predicted to be about $2.4\Msun$ on the causality surface with the equation of state (EOSs) satisfying all known constraints~\citep[e.g.,][]{2019EPJA...55...39Z,2020arXiv200705116L} and it is well known that the centrifugal force in neutron stars rotating at Kepler frequencies can increase their maximum masses by about 20\% with respect to the $M_{\rm TOV}$, the mass of $m_2$ is well within reach with the rotational support. Indeed, this possibility is among the first one considered. However, the conclusions have been diverse mainly
because of the different nuclear EOSs used~\citep[see e.g.,][]{2020ApJ...896L..44A,2020arXiv201002090B,2020arXiv200708493D,2020arXiv200710999G,2020arXiv200804491H,2020arXiv200706057T,2020MNRAS.499L..82M,2020ApJ...902...38Z,2020arXiv200907182Z}. Among the conclusions supporting the $m_2$ as a superfast pulsar, \citet{2020MNRAS.499L..82M} found that $m_2$ has a rotation
frequency of 1210 Hz assuming it has a typical neutron star radius of 12.5 km and a $M_{\rm TOV}$=$2.08\Msun$.
In another study by \citet{2020ApJ...902...38Z}, the $m_2$ was found to have a rotation frequency of 971 Hz using a EOS leading to an equatorial radius of 11.9 km for $m_2$ and a $M_{\rm TOV}$=$2.39\Msun$. More recently, \citet{2020arXiv201002090B} derived a minimum frequency of $1143^{+194}_{-155}$ Hz and an equatorial radius $R_e = 15.7^{+1.0}_{-1.7}$ km for $m_2$ at 90\% confidence level assuming a $M_{\rm TOV}$=$2.14\Msun$ which is the mass of MSR J0740+6620~\citep{2020NatAs...4...72C}.
The minimum frequencies found for $m_2$ in these studies are significantly higher than the fastest known
pulsar PSR J1748-2446ad with a frequency of 716 Hz~\citep{2006Sci...311.1901H}, thus making GW190814's secondary the most massive and fastest known pulsar if confirmed.

While the LIGO/Virgo Collaborations tightly constrained the primary spin of GW190814, the spin of its secondary $m_2$ remains unconstrained~\citep{2020ApJ...896L..44A}. Thus, the possibility for the $m_2$ as a superfast pulsar is neither confirmed nor ruled out observationally. However, an outstanding question critical for the $m_2$ to be a superfast pulsar remains to be answered. Namely, it is unknown if the $m_2$ with the required minimum frequencies are stable against the well-known r-mode instability leading to the exponential growth of gravitational radiation. As it was pointed out already in \citet{2020ApJ...902...38Z,2020arXiv201002090B},
for the $m_2$ to be a supermassive and superfast pulsar, its rotational instabilities would have to be suppressed, otherwise, it is more likely to be a black hole instead.

In this work, we examine the r-mode instability of $m_2$ using well established EOSs satisfying all currently known constraints within the r-mode formalisms established by ~\citet{1998PhRvL..80.4843L,2000PhRvD..62h4030L}.
While a multitude of damping mechanisms exist,
an instability can develop only if its growth is faster than its strongest damping mechanism ~\citep{2020arXiv201002090B}. It is known that a rigid crust provides the strongest r-mode damping, and it can well explain the stability of all observed low-mass x-ray binaries (LMXBs)~\citep[e.g.,][]{2011PhRvL.107j1101H,2012MNRAS.424...93H}. Using this damping mechanism,
we found that the $m_2$ can be r-mode stable as long as its temperature is sufficiently low, e.g., lower than about $3.9\times 10^7 K$ for the $m_2$ rotating at 1169.6 Hz (0.744 times its Kepler frequency). Since this temperature is about an order of magnitude higher than that of some known old neutron stars~\citep[see, e.g.,][]{2019MNRAS.484..974W}, the r-mode instability should not be a concern for the $m_2$ as a supermassive and superfast pulsar within the theoretical framework and models considered.

The r-mode instability~\citep{1998ApJ...502..708A,1998ApJ...502..714F,1998MNRAS.299.1059A,2000ApJ...543..386H} can trigger the exponential growth of gravitational wave (GW) emission in rapidly rotating neutron stars through the Chandrasekhar-Friedman-Schutz mechanism~\citep{1970PhRvL..24..611C,1978ApJ...222..281F} if the GW growth rate is faster than its damping rate. Over the past two decades, significant efforts have been devoted to better understanding the r-mode instability and its damping mechanisms with many interesting findings~\citep[see, e.g.,][]{2000PhRvL..85...10M,2004PhRvD..69h4001S,2005PhRvD..71d4007S,2010PhRvD..82b3007A,2010MNRAS.408.1897H,2011ApJ...735L..29Y,2012PhRvC..85b5801W,2012PhRvC..85d5808V,2012MNRAS.424...93H,2014NuPhA.931..740A,2014PhRvL.113y1102A,2015PhRvD..91b4001I,2015PhRvC..91c5804M,2016ApJ...817..132D,2016Ap&SS.361...98P,2019PhRvD.100j3017O,2019RAA....19...30W,2020PhRvL.125k1106K}. In particular, it was found that the r-mode instability window (a region in the spin frequency versus temperature plane) in which the r-mode is unstable depends sensitively on the EOS of neutron-rich matter, especially its symmetry energy term which encodes the information about the energy cost to make nuclear matter more neutron-rich. It depends even more sensitively on the poorly known properties of neutron star crust, especially whether it is rigid or elastic~\citep[see more discussions in recent reviews, e.g.,][]{2001IJMPD..10..381A,2015IJMPE..2441007H,2017JApA...38...58A,2016EPJA...52...38K}. To realize the goal of this work, it is thus critical to use unified EOSs for neutron stars not only satisfying all currently known constraints from both nuclear physics and astrophysics but also describing the crust, crust-core transition, and the core consistently based on the same nuclear interactions within the same nuclear many-body theories. Moreover, it is important to limit the uncertain r-mode damping mechanisms as much as possible and to identify the strongest one using existing astrophysical observations.

For the purposes outlined above, we describe in the next section how we selected the unified EOSs satisfying the conditions mentioned above. In Section 3, we compare 19 observed LMXBs with known frequencies and temperatures with the r-mode instabilities boundaries calculated with the selected unified EOSs assuming the crust is either perfectly rigid or elastic. As expected, the stability of the 19 LMXBs prefers the perfectly rigid crust.
Assuming the same damping mechanism is at work in other neutron stars, we infer in Section 4 the maximum temperature of $m_2$ by
comparing the minimum frequency it must have to sustain rotationally a mass of $2.50\Msun$ with the maximum frequency it can have before the r-mode instability occurs. We then summarize our work in Sec. 5.

\section{Unified EOSs for neutron stars}
Various EOSs for neutron stars have been used in the literature.
Often, they are constructed from combining the EOSs for the core and crust calculated using different models and/or interactions. For most purposes, these EOSs are perfectly fine.
As we discussed above, for the purpose of this work, it is advantageous to use the EOSs with the crust, crust-core transition, and core all calculated using the same nuclear interaction.
We thus adopt the 33 unified EOSs derived by~\citet{2016PhRvC..94c5804F} within the Skyrme Hartree-Fock and the Relativistic mean-field models for the core and the Thomas–Fermi model for the crust using the same interactions. Because new progress has been made in constraining the EOS since these 33 EOSs were derived, we shall first filter them through the following three tests: (1) satisfying new constraints on the density dependence of nuclear symmetry energy from analyzing both nuclear experiments and astrophysical observations especially the radii and tidal deformability of neutron stars~\citep{2020arXiv200800042N,2020ApJ...899....4X}, (2) being stiff enough to support the presently known maximum mass of neutron stars, i.e., the $M=2.14^{+0.10}_{-0.09}\Msun$ (68\% confidence level) of MSP J0740+6620 \citep{2020NatAs...4...72C}, (3) be consistent with the simultaneous mass and radius measurements by NICER for PSR J0030+0451~\citep{2019ApJ...887L..24M,2019ApJ...887L..22R,2019ApJ...887L..21R}.

\begin{figure}
\vspace{-0.3cm}
\centering
\includegraphics[width=0.45\textwidth]{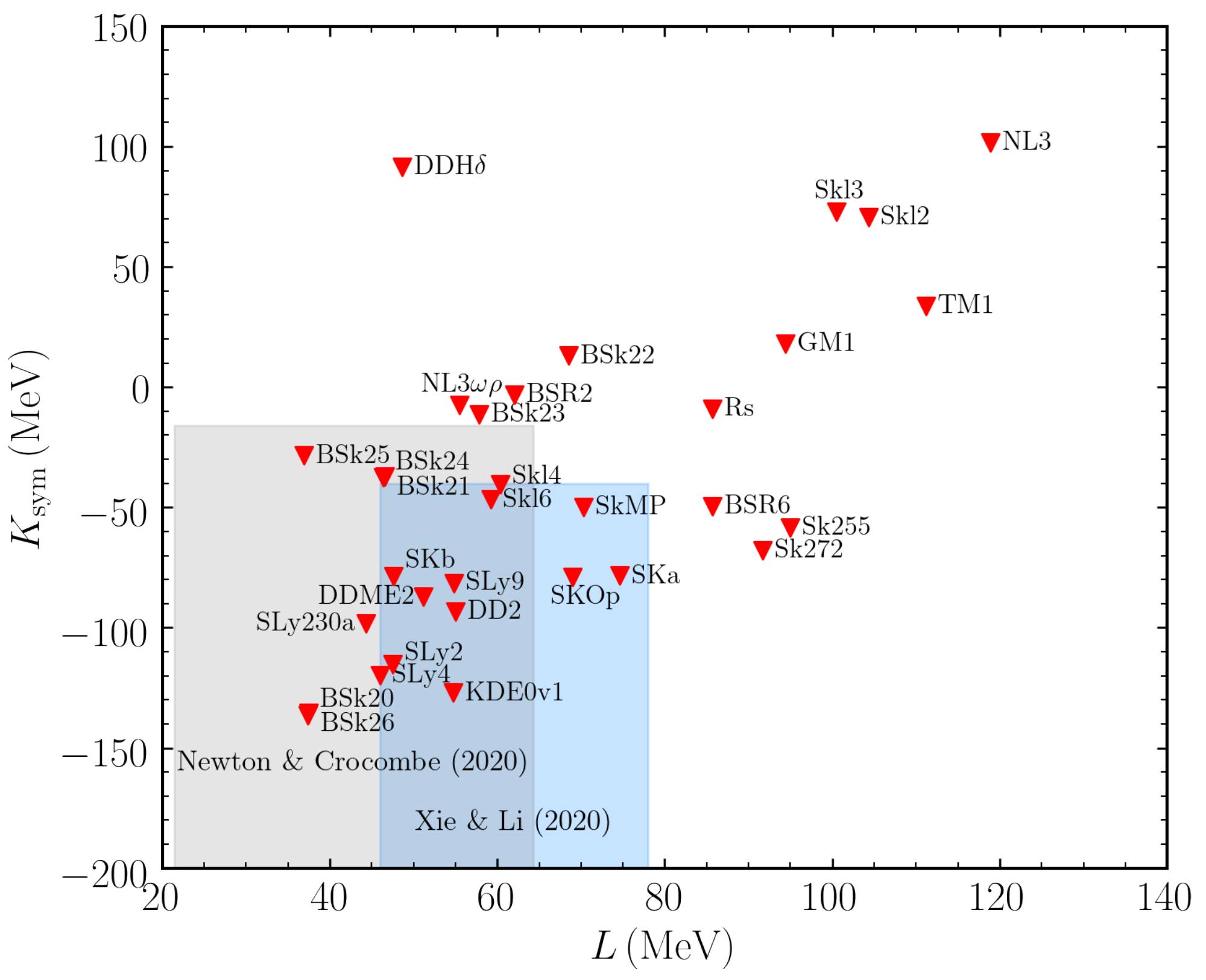}
\caption{$K_{\rm sym}$ parameter as a function of $L$ for the 33 unified neutron star EOSs from ~\citet{2016PhRvC..94c5804F}, together with two recent constraints, to the $68\%$ confidence level, on the slope $L$ and curvature $K_{\rm sym}$ of the symmetry energy at saturation~\citep{2020arXiv200800042N,2020ApJ...899....4X}. Note that the 95\% confidence boundaries shown in \citet{2020arXiv200800042N} are rescaled to the 68\% credible ranges here for a comparison.
Only 7 EOSs: DD2, DDME2, KDE0v1, SKb, SkI6, SLy2, and SLy9 fulfill both constraints.
}
\label{f:ksymL}
\vspace{-0.3cm}
\end{figure}

The nuclear symmetry energy is well-known to have significant effects on the structure of neutron stars \citep{2000PhR...333..121L,2005PhR...411..325S,2019EPJA...55..117L}, including the crust-core transition, the crust thickness, as well as the composition of the star. For a comprehensive review on nuclear symmetry energy and its astrophysical effects, we refer the reader to \citet{2014EPJA...50....9L}.
It is defined as
\begin{eqnarray}
E_{\rm sym} (n) = \frac{1}{2}\frac{\partial^2 E(n,\beta)}{\partial \beta^2}|_{\beta=0}\ ,
\label{eq:sym}
\end{eqnarray}
where $E(n,\beta)$ is the binding energy of neutron-rich matter as a function of density $n$ and isospin asymmetry $\beta = (n_n-n_p)/n$.
The $E_{\rm sym}(n)$ can be characterized by using its slope $L$ and curvature $K_{\rm sym}$ at the nuclear saturation density $n_0$ defined as
\begin{eqnarray}
L&=&3n_0[\partial E_{\rm sym}(n)/\partial n]|_{n=n_0}\ , \\
K_{\rm sym}&=&9n_0^2[\partial^2 E_{\rm sym}(n)/\partial n^2]|_{n=n_0}\ .
\end{eqnarray}

\begin{figure}
\vspace{-0.3cm}
\centering
\includegraphics[width=0.45\textwidth]{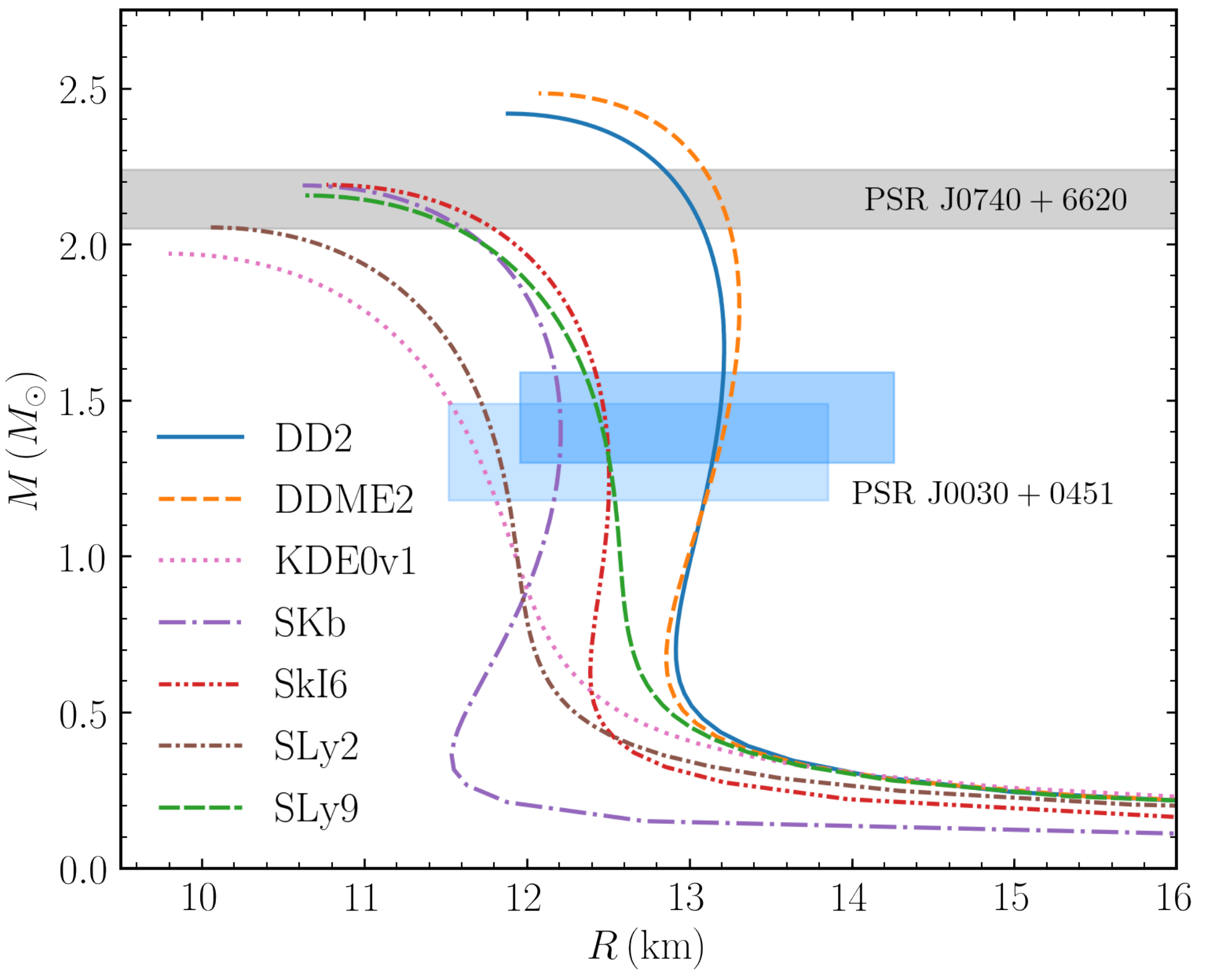}
\caption{Mass-radius relations of neutron stars using the 6 EOSs from Fig.~\ref{f:ksymL} satisfying the latest constraints on $L$ and $K_{\rm sym}$.
The astrophysical constraining bands on the maximum mass~\citep{2020NatAs...4...72C} and the mass-radius correlations for PSR J0030+0451 from NICER~\citep{2019ApJ...887L..24M,2019ApJ...887L..22R,2019ApJ...887L..21R} are shown for comparisons.
}
\label{f:mr}
\vspace{-0.3cm}
\end{figure}

It was shown earlier that both the $L$ and especially $K_{\rm sym}$ as well as their correlation affect the crust-core transition density and pressure significantly~\citep[e.g.,][]{2009ApJ...697.1549X,2009PhRvC..80d5806V,2018ApJ...859...90Z,2019FrASS...6...13P,2020PhRvC.102d5807L}. They also affect the EOS and especially the composition of the core \citep{1991PhRvL..66.2701L}. Fortunately, some significant progresses have been made recently in experimentally constraining the slope parameter $L$~\citep{2013PhLB..727..276L,2016PrPNP..91..203B,2017RvMP...89a5007O}, while the curvature $K_{\rm sym}$ is still much less constrained. In particular, a recent Bayesian analysis of the radii and tidal deformations of canonical neutron stars ~\citep{2020ApJ...899....4X} inferred the most probable values of $L = 66^{+12}_{-20}\,\text{MeV}$ and $K_{\rm sym} = -120^{+80}_{-100}\,\text{MeV}$ at $68\%$ confidence level.
In another Bayesian analysis by
\citet{2020arXiv200800042N} using combined data of neutron skin in $^{48}$Ca, $^{208}$Pb and tin isotopes as well as the best
theoretical information about the EOS of pure neutron matter from {\it ab initio} microscopic nuclear many-body theories, the most probable values of $L$ and $K_{\rm sym}$ were found to be $L = 40^{+34}_{-26}\,\text{MeV}$ and $K_{\rm sym}= -209^{+270}_{-182}\,\text{MeV}$, respectively, at $95\%$ confidence level.

Shown in Fig.~\ref{f:ksymL} are the two constraints on $L$ and $K_{\rm sym}$ in comparison with predictions of the 33 unified EOSs from \citet{2016PhRvC..94c5804F}. It is seen that only 7 of them fall into the overlapping area of the two latest constraints~\citep{2020arXiv200800042N,2020ApJ...899....4X}.
We further test them against the latest astrophysical observations in Fig.~\ref{f:mr}. It is seen that
the KDE0v1 EOS is further excluded by the mass measurement of MSP J0740+6620, to the $68.3\%$ credibility interval, while the remaining 6 EOSs can support a maximum mass of about $2.14\Msun$~\citep{2020NatAs...4...72C}
and satisfy the mass-radius constraints from NICER~\citep{2019ApJ...887L..24M,2019ApJ...887L..21R}.
Thus, we will only use these 6 unified EOSs in our further studies in the following.

\begin{figure*}
\centering
\vspace{-0.3cm}
\includegraphics[width=0.45\textwidth]{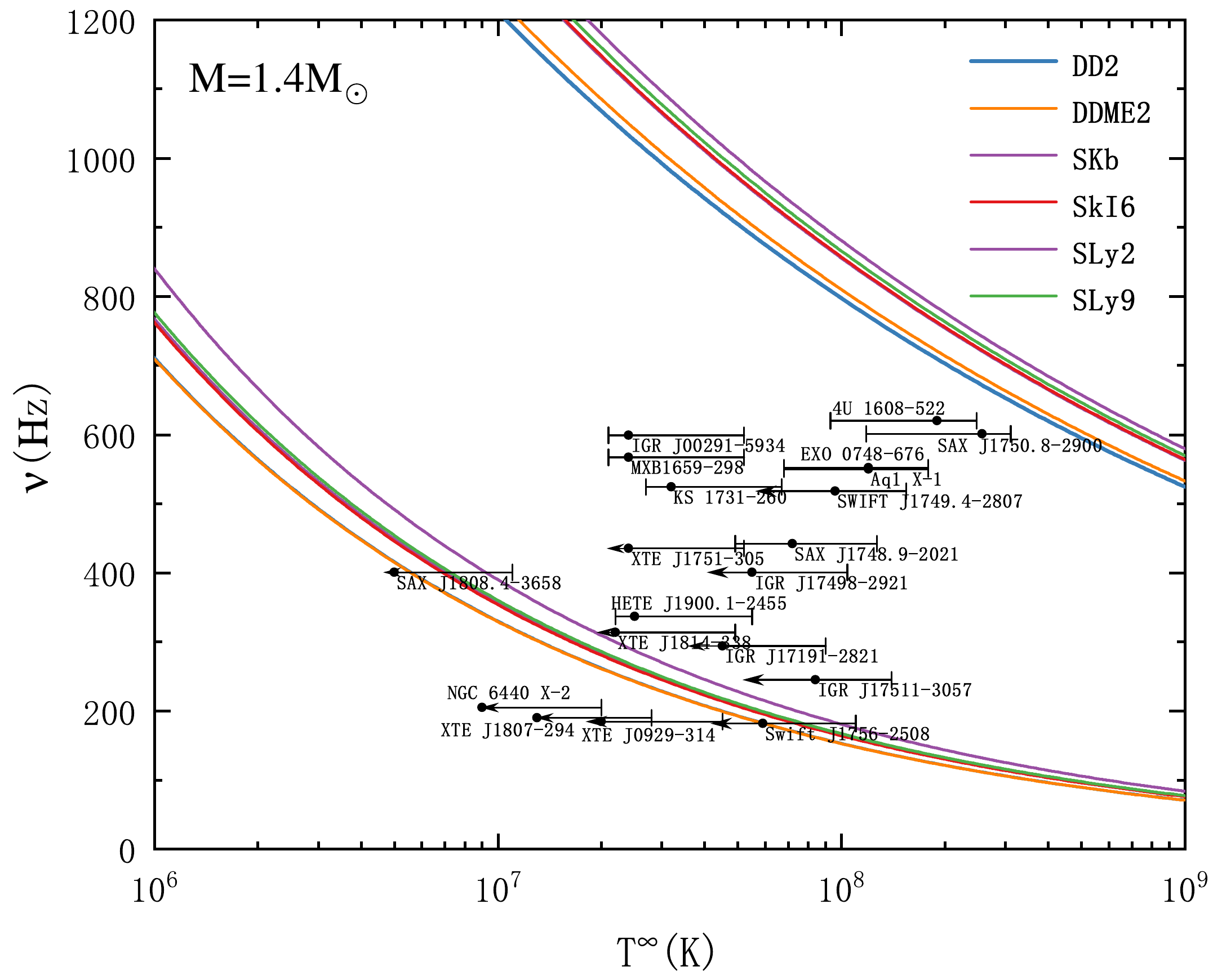}
\includegraphics[width=0.45\textwidth]{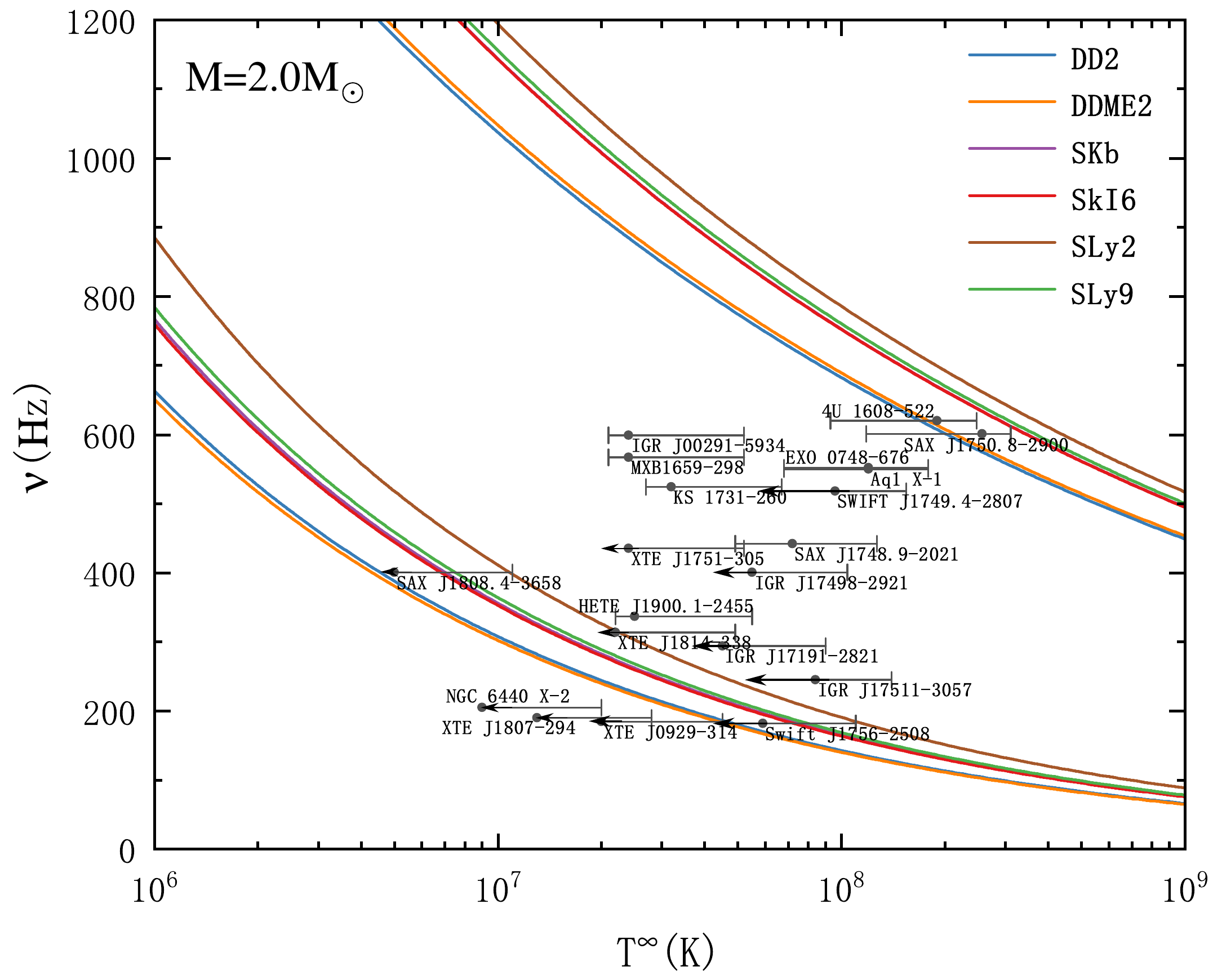}
\caption{Locations of LMXBs in the frequency-temperature plane and lower bounds $\nu_{\rm crit}$ of the r-mode instability windows for $1.4\Msun$ (left) and $2.0\Msun$ (right) neutron stars with 6 selected unified EOSs. The upper (lower) group of lines are calculated with rigid (elastic) crusts.
The observational data and internal redshifted temperatures ($T^{\infty}=T\sqrt{1-2GM/Rc^2}$) on neutron stars in LMXBs are taken from \citet{2014PhRvD..90f3001G, 2017MNRAS.468..291C}, with the error bars due to the uncertainty in the composition of the envelope~\citep[see detail in][]{2014PhRvD..90f3001G, 2017MNRAS.468..291C}.
}
\label{f:rwithdata}
\end{figure*}
\section{R-mode instability in neutron stars}
All rotating neutron stars are generically r-mode unstable due to the emission of gravitational waves. However, there are several possible damping mechanisms preventing the r-mode oscillation from growing exponentially. Here we adopt two extreme damping models assuming the crust is either rigid or elastic.
For easy of discussions, in the following, we first recall briefly some of the basic definitions and formulas most relevant for our study about the r-mode instability of $m_2$. We refer the reader to the original papers for more details.
\subsection{Time scales of r-mode growth and damping}
Whether the r-mode oscillation can grow exponentially or not
depends on the competition between the growth rate of gravitational wave emission and the r-mode damping rate (mainly due to viscosity). Thus, the r-mode instability boundary is determined by setting the damping $\tau_{\rm diss}$ and driving $\tau_{\rm gw}$ timescales equal to each other~\citep{2001IJMPD..10..381A}, i.e., by solving for the zeros of
\begin{equation}
      -\frac{1}{2\tilde{E}}\left(\frac{d\tilde{E}}{dt}\right)=\frac{1}{\tau_{\rm gw}(\nu)}+\sum\frac{1}{\tau_{\rm diss}(\nu,T)}=0 \ ,
      \label{eq:window}
\end{equation}
where $\tilde{E}$ is the total energy of an r-mode oscillation. It is known that the shear viscosity dominates the energy dissipation for temperatures around $10^8~\rm{K}$, while at higher temperatures (of the order of $10^{10}$ K) the bulk viscosity becomes important~\citep{2001IJMPD..10..381A}.
Since all the observed LMXBs
have temperatures around $10^8~\rm{K}$ and the old neutron stars are cooler, in the present study we consider only the shear viscosity resulting from both neutron-neutron (nn) and electron-electron (ee) scatterings. We adopt the following parameterizations for the two viscosities as functions of temperature $T$~\citep{1979ApJ...230..847F,1987ApJ...314..234C,2008PhRvD..78f3006S},
\begin{eqnarray}
\eta_{nn}(r) &=& 347 [\rho(r)] ^{9/4} T^{-2}~~\rm{(g~cm^{-1} s^{-1})} \ ,\\
\eta_{ee}(r) &=& 6\times 10^6 [\rho(r)]^2 T^{-2}~~\rm{(g~cm^{-1} s^{-1})}\ ,
\end{eqnarray}
where $\rho(r)$ is the mass density profile of the star.

The crust is expected to play an important role in determining the r-mode stability of neutron stars. Here we consider two extreme cases. Assume the crust is perfectly rigid (i.e., the ``slippage'' factor $S = 1$), so the fluid motion associated with the r-mode would be significantly damped at the crust-core boundary, the $\tau_{\rm gw}$ is given by~\citep{2000PhRvD..62h4030L},
\begin{eqnarray}
   & \frac{1}{\tau_{\rm gw}^{
    \rm c}} = -\frac{32\pi G \Omega^{2l+2}}{c^{2l+3}} \frac{(l-1)^{2l}}{[(2l+1)!!]^2}\left(
    \frac{l+2}{l+1}\right)^{2l+2}       \nonumber\\
    &\times \int_{0}^{R_{\rm c}} \rho(r) r^{2l+2} dr \ ,
    \label{eq:taug_c}
\end{eqnarray}
where $G$ is the gravitational constant, $\Omega=2\pi\nu$ is the angular velocity of the star, $c$ is the speed of light and $R_c$ is the stellar radius at the crust-core transition mass density $\rho_c$. While the $\tau_{\rm diss}$ can be evaluated from~\citep{2000PhRvD..62h4030L},
\begin{equation}
\tau_{\eta}^{\rm c} = \frac{1}{2\Omega}\frac{2^{l+3/2}(l+1)!}{l(2l+1)!!\mathcal{C}_l}\sqrt{\frac{2
\Omega R_c^2 \rho_c}{\eta_c}}\int_0^{R_c}\frac{\rho}{\rho_c}\left(\frac{r}{R_c}\right)^{2l+2}\frac{dr}{R_c},
\label{eq:taue_c}
\end{equation}
where $\eta_c$ is the viscosity at the crust-core transition density.
Following \citet{1998PhRvL..80.4843L}, we only consider the lowest order contribution of the 
multipole moment ($l = 2$), with $\mathcal{C}_2=0.80411$~\citep{2001ApJ...550..443R}.

In the other extreme case neglecting the crustal damping, or
the crust is regarded as extremely elastic [i.e., the ``slippage'' factor $S\rightarrow0$~\citep[see e.g.,][]{2006PhRvD..74d4040G}], the crust will also participate in the oscillation. In this case, the $\tau_{\rm gw}$ is given by~\citep{1998PhRvL..80.4843L},
\begin{eqnarray}
   & \frac{1}{\tau_{\rm gw}} = -\frac{32\pi G \Omega^{2l+2}}{c^{2l+3}} \frac{(l-1)^{2l}}{[(2l+1)!!]^2}\left(
    \frac{l+2}{l+1}\right)^{2l+2}   \nonumber \\
   &\times \int_{0}^{R} \rho(r) r^{2l+2} dr \ .
    \label{eq:taug}
\end{eqnarray}
While the $\tau_{\eta}$ is written as~\citep[e.g.,][]{2012PhRvC..85d5808V},
\begin{eqnarray}
    &\frac{1}{\tau_{\eta}} = (l-1)(2l+1)\left(\int_0^{R}\rho(r) r^{2l+2} dr\right)^{-1}     \label{woc}     \label{eq:eta}    \nonumber \\
    &\times \int_{0}^{R} \eta(r) r^{2l} dr\ .
        \label{eq:taue}
\end{eqnarray}

The actual neutron star crust should have some elasticity, therefore by calculating the r-mode boundary in the two extreme cases: Eqs.~(\ref{eq:taug_c})-(\ref{eq:taue_c}) (where a rigid crust is considered) and Eqs.~(\ref{eq:taug})-(\ref{eq:taue}) (where no crustal damping is included), we may estimate the uncertainty due to the crustal modelling. Moreover, confronting the theoretical results with the observations of neutron stars' frequency and temperatures in LMXBs, we may get some hints about which damping mechanism dominates.

\subsection{Lower boundaries of r-mode instability windows}
From Eq.~(\ref{eq:window}), for each star at a given temperature, one can find its critical frequency $\nu_{\rm crit}$ above which the star becomes unstable against the running-off of gravitational radiation, by setting $1/{\tau_{\rm gw}(\nu_{\rm crit})}+1/\tau_{\eta}(\nu_{\rm crit},T)=0$.
The $\nu_{\rm crit}$ serves as an upper limit of stable pulsars' frequencies for a given temperature. The region above this boundary in the frequency-temperature plane is the so-called r-mode instability window. In the following, we examine the r-mode instability boundaries with the 6 selected EOSs with respect to the locations of the 19 LMXBs. Since the masses of the neutron stars in these LMXBs are not measured accurately, the calculations are done for both canonical ($M=1.4\Msun$) and massive ($M=2.0\Msun$) neutron stars.

Our results are shown in Fig. \ref{f:rwithdata} for both cases assuming a perfectly rigid crust (upper groups of lines) or an elastic one (lower group of lines). Several interesting observations can be made:
\begin{itemize}
\item The r-mode instability boundary depends most sensitively on the crust's elasticity while effects of the EOS and neutron star mass are appreciable. Most importantly, for the present study, the rigid crust provides the strongest damping.

\item The damping mechanism with a perfectly rigid crust can accommodate all neutron stars in the 19 LMXBs, while the one with an elastic crust can only accommodate a few slowly-rotating neutron stars, such as NGC 6440 X-2, XTE J1807-294, and XTE J0929-314.

\item The r-mode instability boundaries are sensitive to the EOS especially the $L$ parameter~\citep[see e.g.,][]{2012PhRvC..85b5801W}, but the effects shown here are much less than previously reported in the literature mainly because the 6 unified EOSs used here are much more stringently constrained compared to those used before.
Nevertheless, one can still clearly see that the $L$ parameter (shown in Fig.~\ref{f:ksymL}) can not be very large, e.g., smaller than $\sim60\,\text{MeV}$ if LMXB neutron stars are massive (e.g., $M=2.0M\Msun$), to ensure the rapidly-rotating neutron stars in, e.g., 4U 1608-522 and SAX J1705.8-2900, stay r-mode stable.

\item
Comparing the results for $1.4\Msun$ and $2.0\Msun$ neutron stars, we see that the r-mode instability window is generally broader for the latter in agreement with earlier findings~\citep[e.g.,][]{2012PhRvC..85d5808V}, especially in the case with the rigid crust. Moreover, the uncertainty band due to the EOS is relatively wider for neutron stars of mass $2.0\Msun$, due to the more significant diversity of the EOS at higher densities reached.
\end{itemize}

\begin{figure}
\vspace{0.3cm}
\centering
\includegraphics[width=0.45\textwidth]{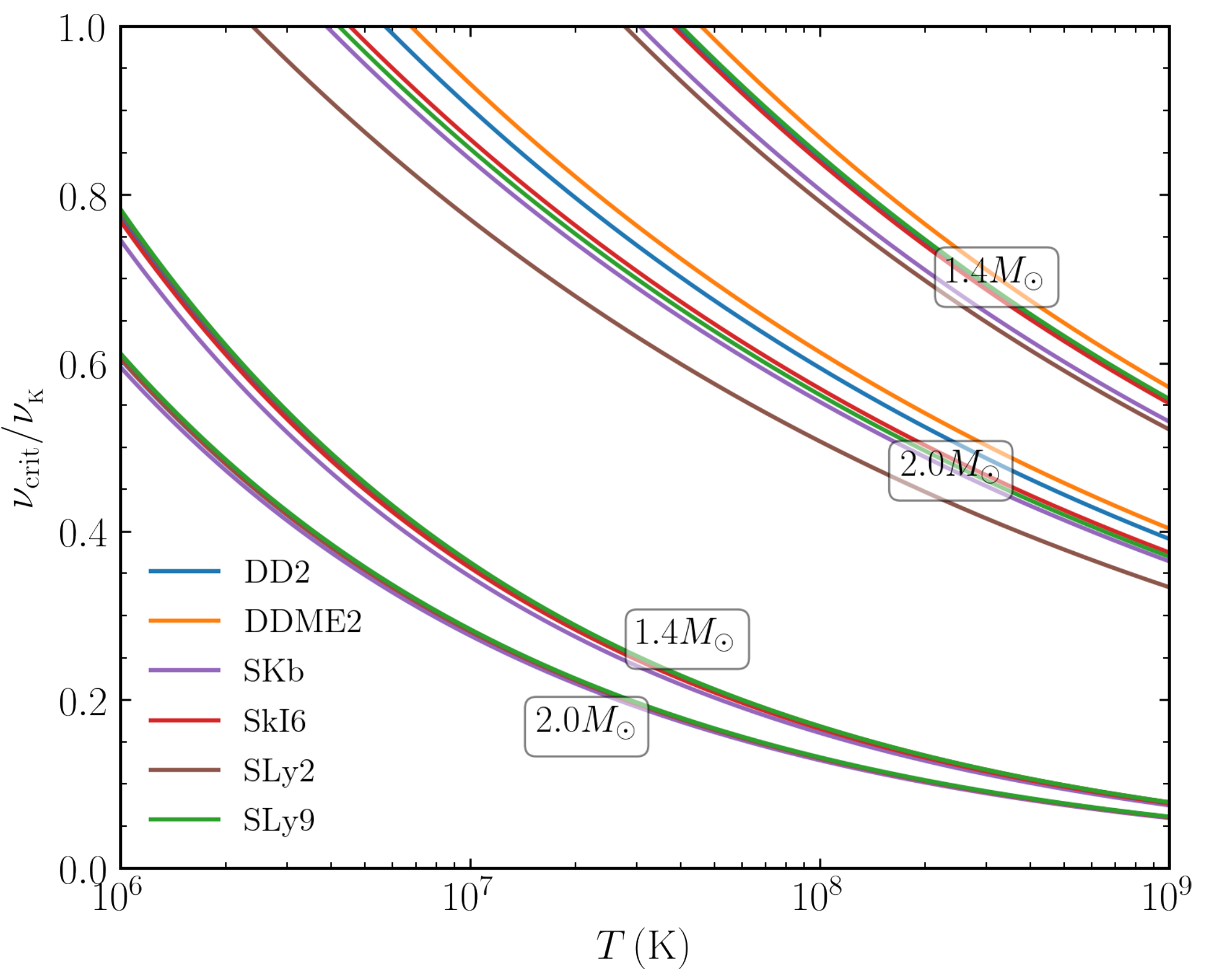}
\caption{Temperature dependence of the reduced critical frequency $\nu_{\rm crit}/\nu_{\rm K}$ for $M=1.4\Msun$ and $M=2.0\Msun$ neutron stars with rigid (upper groups of lines) or elastic crusts (lower groups of lines).
}
\label{f:vkT}
\end{figure}

Since the Kepler frequency $\nu_{\rm K}$ is the absolute upper limit of spin frequencies of all stars with a given mass, it is interesting and informative to examine the reduced critical frequency $\nu_{\rm crit}/\nu_{\rm K}$ as functions of the temperature $T$ for both $1.4\Msun$ and $2.0\Msun$ neutron stars. Our results for $\nu_{\rm crit}/\nu_{\rm K}$ are shown in Fig. \ref{f:vkT}. Compared to the un-scaled results shown in Fig.~\ref{f:rwithdata}, it is seen that the r-mode instability boundaries of the reduced frequency are more sensitive to the neutron star mass with both rigid and elastic crusts. This is because the $\nu_{\rm K}$ is proportional to the square root of the average density of neutron stars~\citep{2001IJMPD..10..381A}.
The average density depends on the mass; thus, the reduced frequency becomes more sensitive to the neutron star mass. The lighter neutron stars have a lower average density and a lower $\nu_{\rm K}$, making them have higher reduced frequencies than the massive ones.

Since the Kepler frequency is EOS dependent~\citep[see, e.g.,][]{2016PhRvD..94h3010L,2017ApJ...844...41L}, the reduced frequencies thus obtain a more complicated EOS dependence. With the rigid crust, as shown in
Eqs.~(\ref{eq:taug_c})-(\ref{eq:taue_c}), the location $R_c$ of the crust-core transition plays an important role. In this case, the strong EOS dependence of $R_c$ makes the reduced frequency more strongly EOS dependent than the case with an elastic crust. While in the latter case, both the
$\nu_{\rm crit}$ and $\nu_{\rm K}$ are calculated by integrating from the center to the surface. The reduced frequency obtained appears rather EOS insensitive due to some canceling effects in taking the ratio. Finally, the difference in the reduced frequency between the two cases at the same temperature are very large. More quantitatively, for a typical temperature $T=10^8~\rm K$ relevant for the LMXBs, $\nu_{\rm crit}$ is lifted from $\sim0.16~\nu_{\rm K}$ ($\sim0.13~\nu_{\rm K}$) to $\sim0.79~\nu_{\rm K}$ ($\sim0.48~\nu_{\rm K}$) for $M=1.4\Msun$ ($M=2.0\Msun$) neutron stars when assuming a rigid crust with respect to the elastic one.

In short, the main lessons we learned from this section that are most important for our following study about the r-mode stability of $m_2$ are (1) the rigid crust provides the strongest damping, (2) the stability of the 19 neutron stars in LMXBs favors the rigid crust damping, (3) the reduced critical frequency $\nu_{\rm crit}/\nu_{\rm K}$ with the rigid crust is sensitive to both the neutron star mass and the EOS used.

\section{R-mode stability of GW190814's secondary component of mass $2.50\Msun$}

We now turn to examine the r-mode stability of GW190814's secondary component $m_2$ as a supermassive and superfast pulsar. For this purpose, we first find the minimum frequency $\nu_{\rm min}$ to support GW190814's secondary component with a minimum mass of $2.50\Msun$~\citep{2020ApJ...896L..44A} using the well-tested ${\rm RNS}$ code \citep{RNS} for a given EOS. Among the 6 EOSs used above, we found that the SLy2 EOS can not rotationally support a $2.50\Msun$ star even at the Kepler frequency. Thus, only the remaining 5 EOSs are used. As discussed in the previous section, only the damping with the rigid crust is necessary for this discussion as it provides the strongest/quickest damping.

\begin{table}
\caption{\small Neutron star properties based on the 5 unified EOSs models~\citep{2016PhRvC..94c5804F}. $M_{\rm TOV}$ is the maximum mass of non-rotating neutron stars supported by the given EOS. $\nu_{\rm min}$ is the minimum frequency (in unit of the Kepler frequency $\nu_{\rm K}$) to support GW190814's secondary component $m_2$ with a minimum mass of $2.50\Msun$.
$T_{\rm max}$ is the maximum temperature for the $m_2$ to remain r-mode stable.
}
\setlength{\tabcolsep}{1.1pt}
\renewcommand\arraystretch{0.95}
\begin{ruledtabular}
\center\begin{tabular}{ccccc}
Model&$M_{\rm TOV}$&$\nu_{\rm K}$ & $\nu_{\rm min}$ & $T_{\rm max}$\\
&$(\Msun)$&(Hz)& ($\nu_{\rm K}$) & (K) \\
 \hline
DD2  & $2.42$  & $1196.6$  &$0.755$ &$3.0\times 10^7$\\
DDME2 & $2.48$ & $1169.6$  & $0.744$ & $3.9\times 10^7$\\
SKb  &$2.20$  &$1446.5$  & $0.809$ & $1.3\times 10^7$\\
SkI6 & $2.20$  & $1433.4$  & $0.827$ & $1.1\times 10^7$\\
SLy9 & $2.16$  & $1515.4$  & $0.855$ & $7.5\times 10^6$\\
\end{tabular}
\label{tab:vmin}
\end{ruledtabular}
\end{table}

\begin{figure}
\vspace{0.3cm}
\centering
\includegraphics[width=0.45\textwidth]{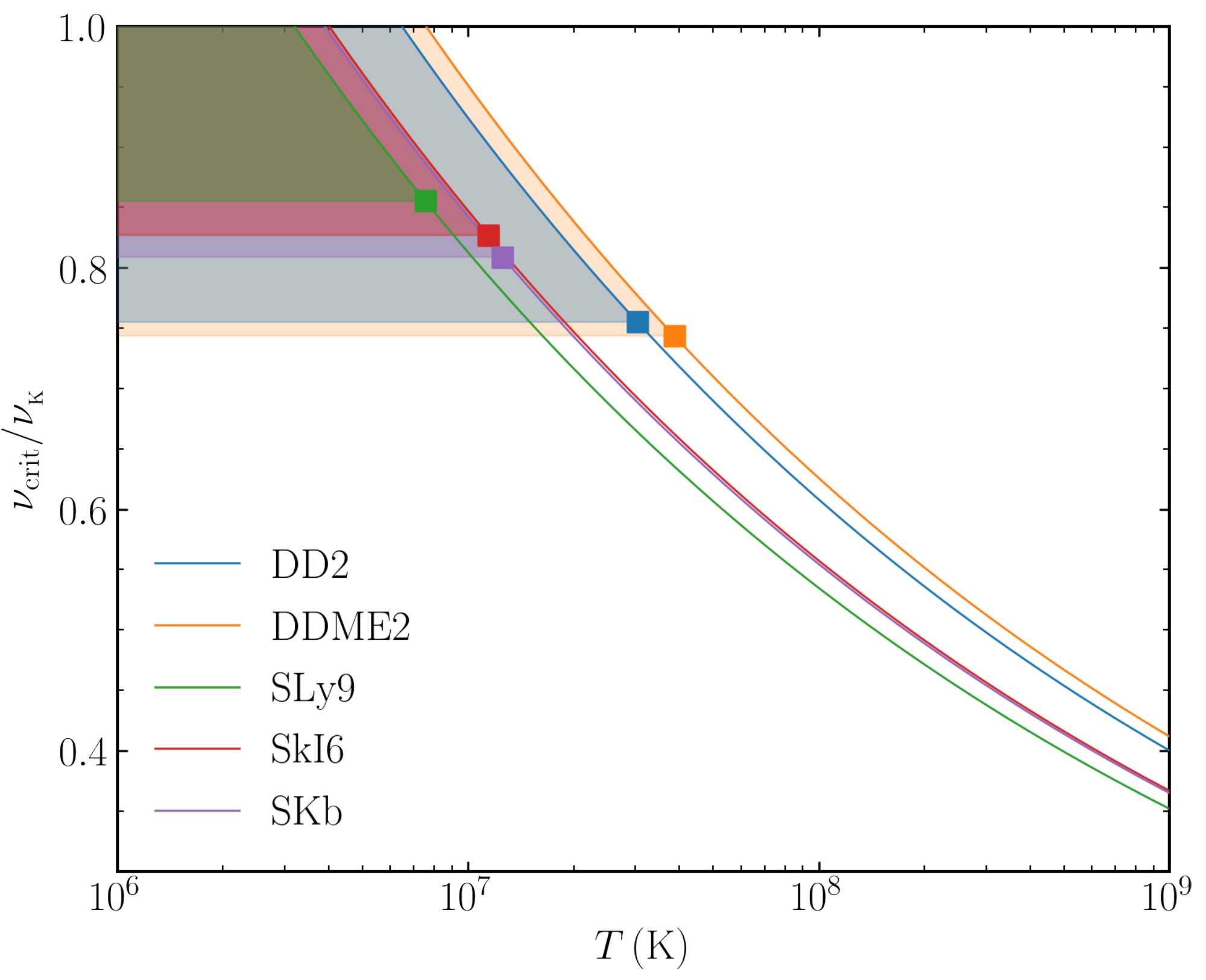}
\caption{Temperature dependence of the reduced critical frequency $\nu_{\rm crit}/\nu_{\rm K}$ for neutron stars with rigid crusts. For each EOS, the $\nu-T$ space allowing the secondary component of GW190814 as a r-mode stable pulsar is shown as the shaded region. The maximum temperatures $T_{\rm max}$ below which neutron stars remain r-mode stable are marked with diamonds on the r-mode instability thresholds for the 5 EOSs used.
}
\label{f:vkT25}
\end{figure}

Listed in Table 1 are the maximum mass of non-rotating neutron stars $M_{\rm TOV}$, Kepler frequency $\nu_{\rm K}$ and the minimum frequency $\nu_{\rm min}$ to support a neutron star of mass $2.50\Msun$ for the 5 EOSs used. The $\nu_{\rm min}$ is then plotted as a horizontal line in Fig. \ref{f:vkT25} where the r-mode stability boundary for each EOS is shown in the frequency-temperature plane. Their cross point is marked with a diamond indicating the maximum temperatures $T_{\rm max}$ below which neutron stars remain r-mode stable. The resulting 5 maximum temperatures are listed in the last column in Table 1. While the allowed $\nu-T$ parameter spaces are reported with the shaded regions in Fig.~\ref{f:vkT25}.

As expected, the $\nu_{\rm min}$ depends sensitively on the
$M_{\rm TOV}$ determined by the stiffness of the EOS.
For the 5 EOSs used, the $\nu_{\rm min}$ and $M_{\rm TOV}$ range between 0.744 to $0.855\,\nu_{\rm K}$ and 2.16 to $2.42\Msun$, respectively. The corresponding maximum temperature $T_{\rm max}$ is between $7.5\times 10^6$ to $3.9\times 10^7$ K. This temperature limit is about 10 times cooler than the LMXBs but also
about 10 times hotter than some of the known old neutron stars~\citep{2019MNRAS.484..974W}. This information may be useful for further understanding the dynamical path of the GW190814 event and the nature of its secondary component.

The required minimum frequencies $\nu_{\rm min}$ between 870 Hz and 1295 Hz found here with the 5 selected EOSs satisfying all currently known constraints from both nuclear physics and astrophysics cover the minimum frequencies found necessary for the $m_2$ to rotationally sustain a minimum mass of $2.50\Msun$ in~\citet{2020MNRAS.499L..82M,2020ApJ...902...38Z,2020arXiv201002090B}.
As discussed above, as long as the $m_2$ has a temperature below about $3.9\times 10^7$ K, the $m_2$ spinning with the high frequencies found above can remain r-mode stable.
While it is not the purpose of this work to study how the high mass and spin can be obtained, it is very interesting to note that it was already pointed out by \citet{2020ApJ...899L..15S} that they could be supplied through the circumbinary accretion disk if the $m_2$ was born as a neutron star where a significant amount of the supernova ejecta mass from its formation remained bound to the binary due to the presence of the massive black hole companion in the GW190814 event.

\section{Conclusions}
In conclusion, the GW190814's secondary component $m_2$ can be an r-mode stable supermassive and superfast pulsar rotating with a frequency higher than 870.2 Hz (0.744 times its Kepler frequency of 1169.6 Hz) as long as its temperate is lower than about $3.9\times 10^7 K$ which is still about 10 times hotter than some of the known old neutron stars. Thus, the r-mode instability should not be a concern for the $m_2$. The minimum frequency and limiting temperature found here may be useful for further understanding the formation mechanism of $m_2$ and the merger dynamics of GW190814.

Our study is carried out within the minimum model of neutron stars consisting of only nucleons and leptons. Moreover, the fast rotation is probably the simplest mechanism to provide the additional support besides the nuclear pressure against the strong gravity of massive neutron stars. Furthermore, we selected the EOSs satisfying all currently known astrophysical and nuclear physics constraints from a group of 33 unified EOSs constructed from the crust to the core consistently using the same nuclear interactions. Thus, our approach is probably among the most conservative ones used in the literature in investigating the nature of GW190814's secondary component.

A major caveat of our work is the assumption that the rigid crust of neutron stars provides the strongest r-mode damping among a multitude of possible damping mechanisms. While both our current work and previous theoretical calculations by others as well as the stability of the 19 neutron stars in LMXBs favor the dominating rigid crust damping, there may be other even stronger damping mechanisms and/or model ingredients affecting significantly the location of the r-mode instability windows. Nevertheless, our current study can successfully describe all relevant observations of neutron stars. We are thus confident that our present conclusions are physically sound and useful for the community to finally solve the mystery regarding the nature of GW190814's secondary component.

\acknowledgments
We would like to thank Morgane Fortin for valuable discussions.
This work is supported in part by National SKA Program of China (No. 2020SKA0120300), the National Natural Science Foundation of China (Grant Nos. 11873040, 12033001), the Youth Innovation Fund of Xiamen (No. 3502Z20206061), the CAS ``Light of West China'' Program (No. 2019-XBQNXZ-B-016), the Tianshan Youth Program (No. 2018Q039), the U.S. Department of Energy, Office of Science, under Award Number DE-SC0013702, the CUSTIPEN (China-U.S. Theory Institute for Physics with Exotic Nuclei) under the US Department of Energy Grant No. DE-SC0009971.

\newpage

\end{document}